# CAN Networks Security in Smart Grids Communication Technologies

Ayman W. Baharia[1], Khaled T. Naga[2], Hesham S. Abdelfattah[3], Shady A. Maged[4], Sherif A. Hammad[5]

[1]Mechatronics Department, Faculty of Engineering, Ain Shams University, Cairo, Egypt. ayman_baharia@eng.asu.edu.eg
[2]Automotive Cybersecurity Expert, Cairo, Egypt. eng.khalednaga@gmail.com
[3]Mechatronics Department, Faculty of Engineering, Ain Shams University, Cairo, Egypt. heshammousa95@eng.asu.edu.eg
[4] Mechatronics Department, Faculty of Engineering, Ain Shams University, Cairo, Egypt. shady.maged@eng.asu.edu.eg
[5]Garraio for Software Innovations, Cairo, Egypt. sherif.hammad@garraio.com

***ABSTRACT:*** *The rapid evolution of smart grids requires effective communication protocols to transfer data reliably and securely. Controller Area Network (CAN) is one of the most recognized protocols that offer reliable data transmission in smart grids due to its robustness, real-time capabilities, and relatively low initial cost of its required hardware. However, as a smart city becomes more interconnected, it also becomes more vulnerable to cyber-attacks. As there are many mechanisms to secure the CAN nodes from attacks, most of those mechanisms have computational overhead, resulting in more delay in the network. We implemented a solution that requires almost no overhead to any CAN node connected to the network. It depends on a single node responsible for securing the CAN network. This approach seeks to augment network security while reducing security mechanisms overhead to all CAN network nodes. The methodology and comprehensive test results will be presented in detail during a subsequent discussion. The used software for development is Code Composer Studio, and the used microcontroller evaluation boards (EVB) are TM4C 1294.*

## 1. INTRODUCTION

Modern smart grids rely on network communication protocols to provide efficient data exchange and reliable grid management. Communication networks are at risk because they can be easily accessed without authentication. This is primarily because communication between the Electronic Control Units (ECUs) happens without any built-in authentication, leaving them exposed to unauthorized access.

### 1.1. Literature Review
#### 1.1.1. CAN Protocol

Controller Area Network (CAN) operates as a multi-master system where messages are broadcasted to all connected nodes. Only the one with the intended message ID will process the frame [1]. It supports two operational standard and extended modes. Standard mode has an 11-bit message identifier. Whereas the Extended mode supports 29-bit message identifier. Various types of CAN frames exist, including data and remote frames for data transfer, error frames for reporting errors, and overload frames for managing flow control. Additionally, CAN offers high noise resistance due to the use of differential twisted-pair lines for data transmission [2].

CAN network is vulnerable to attacks by hackers who can gain access and control to the grid. As the CAN protocol itself does not provide strong security mechanisms, securing the CAN node is a must during the early phases of smart grids development.

#### 1.1.2. CAN Protocol Applications

Nowadays, CAN networks have wide applications in smart micro/large grids, real-time monitoring and control, and the automotive industry [3].

#### 1.1.3. Attacks on CAN Networks

As a CAN network lacks security and authenticity, many researchers tested different attacks on CAN networks.

Research studies implemented a "Fuzzing" testing, where an attacker CAN node transmits random CAN frames and waits for a system response. This method does a sort of reverse engineering to take control over a network [4].

Verdult et al. succeeded to attack and access a CAN network through direct connection [5].

### 1.1.4. RbT Protocol

K. Naga compared different security mechanisms implemented in CAN network. Busload overheads are shown when implementing security mechanisms. It is even higher when the CAN node tries to recover from an attack [1]. Then, Rule-based Transceiver (RbT) node is introduced. The transceiver acts as a monitoring ECU that samples and analyses the CAN frame in real-time while still being transmitted on the CAN bus. It detects malicious frames and injects a CAN error frame during their transmission (just after discovering a malicious attack in any CAN frame section). This destroys the CAN frame sent by an attacker, preventing it from reaching the receiver CAN node that was meant to be attacked.

This "Firewall/RbT" node defers software overhead from all CAN nodes to its protocol.

### 1.1.5. Techniques Implemented to Secure CAN Networks

Hartkopp et al. introduced a protocol called "MaCAN" that uses time-based mechanisms to enhance CAN network security. It includes a time server hardware node to add global timing through timestamps and a key server hardware node to manage key distribution. The protocol uses a 4-byte Message Authentication Code (MAC) for security. While this method is efficient, it takes up 50% of the message payload, causing significant overhead and increasing busload. The large size of the MAC may also require splitting the message across multiple frames. Additionally, there could be challenges during the transmission of synchronization messages [6].

Moreover, M. Mostafa et al. proposed a technique using specific hardware components that goes deep into bit-stream level testing and error injection [7]. It requires hardware modifications in the microcontroller.

Furthermore, Radu et al. defined a lightweight protocol "LeiA" that provides strong authentication. This protocol is based on sending a separate frame for MAC message but imposes high busload overhead [8]. Shashwat el al. proposed a solution that A lightweight, FPGA-based IDS architecture that monitors CAN bus communication in real time [9]. Other techniques were proposed by researchers to address the increasing threats of CAN networks cyberattaks [10].

The mentioned techniques rely on having the algorithms overhead on the CAN node itself. This paper aims at minimizing this overhead without any hardware modifications.

### 1.2. Study Objectives

This paper aims at designing and implementation of malicious CAN frames detection/prevention solution. Off-the-shelf evaluation board is the main platform. Besides, CAN bus desktop monitoring tool is implemented to verify the achievement.

### 1.3. Contributions

Research outcome shows an affordable software/hardware cybersecurity enhancement on CAN networks.

### 1.4. Paper Organization

The remainder of this paper will be organized as follows: the literature review, including the communication protocol, the recent attacks and the security techniques, the research methodology, the results, the conclusion, and the future work.

## 2. RESEARCH METHODOLOGY

The first step is sampling the CAN frame being transmitted on the CAN bus. This is done by a separate ECU (Firewall/RbT node). The RbT node is responsible of analysing the frame, and acting based on a certain cyber security algorithm. The main objective is to prevent malicious CAN frame from reaching the target CAN node [1]. CAN/Ethernet send/receive monitoring tool is typically designed and implemented for this purpose. Its developed firmware and desktop libraries are financed by ITIDA [11] and freely available for education purpose.

## 3. RESULTS

Experimental tests are done to each implementation phase. Typically, they are CAN frame sampling, error frame generation and injection, and killing the attacker CAN frame and turning the attacker CAN node into Bus-Off state.

### 3.1. Experimental Setup

As shown in figure 1, CAN 'Transceiver0' RX/TX are connected to GPIO pins of the RbT node for sampling and error frame injection. TM4C 1294 EVB is used as RbT node. This allows bit by bit checking and injection during CAN frame transmission on the bus. CAN monitoring desktop tool is used to monitor and attack the network. It has Ethernet communication with TM4C 1294 EVB that is physically connected to the CAN network.

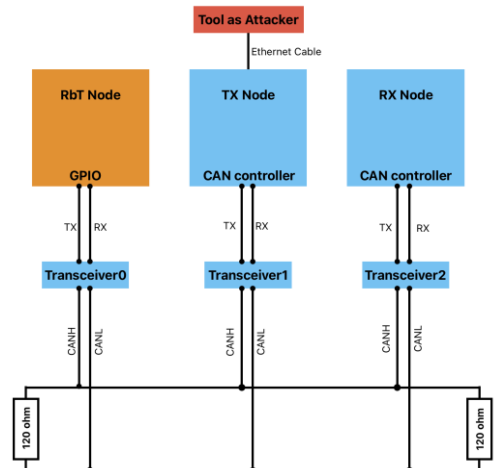

*Figure 1: Experimental setup.*

### 3.2. Software Monitoring and Controlling Tool

A desktop sender/receiver tool, shown below in figure 2, is developed. CAN frames are configurable (e.g., name, ID, and data).

The tool also offers creating ethernet frames, searching, importing/exporting frames from/to excel sheets. The generated CAN frames are used to test the network security. Many frames are created with different identifiers. These frames are sent from the desktop tool to the TX node through ethernet RJ45 cable. The valid frames are sent on the bus normally. The attacker frame is detected/prevented by the RbT node. Therefore, the attacking node couldn't send frames anymore.

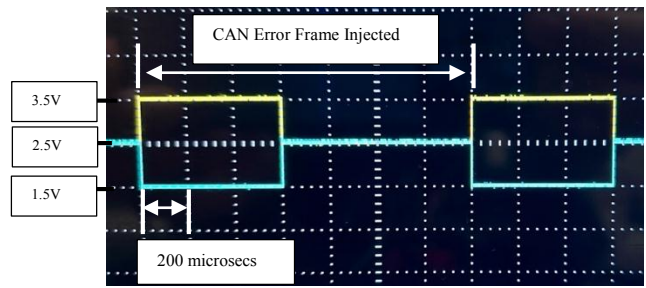

*Figure 2: Desktop Sender/ Receiver tool.*

### 3.3. Sampling the CAN Frame

To successfully sample the CAN frame, TM4C1294 EVB is used as RbT node [12]. Composer Studio (CCS) is the IDE [13]. Sampling is done using a CAN transceiver connected between the CAN bus and the RbT node, having the raw CAN frame stored/analysed. Sampling a CAN frame is validated on an oscilloscope as shown in figure 3. The sampled frame is successfully stored and processed for removal of all stuff bits and recognition of Start Bit, ID, RTR, IDE, R0, DLC, and data. Typical CAN network communication is at 10 Kbps bitrate.

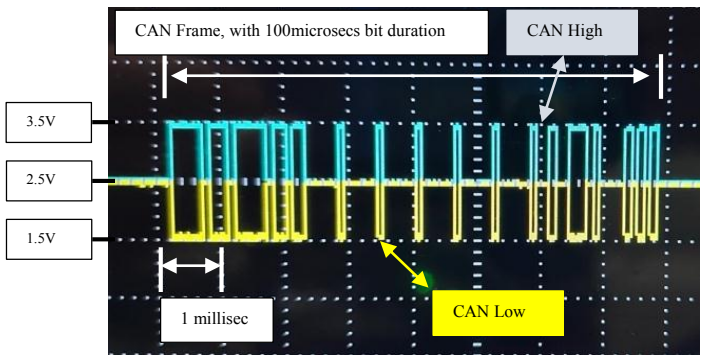

*Figure 3: CAN frame validation on oscilloscope.*

### 3.4. RbT Algorithm

All nodes register their messages IDs initially, then any frame having an ID that is not registered will be treated as a malicious frame sent by an attacker. The malicious frame will be killed, and the attacker CAN node state will be Bus-Off.

### 3.5. Generating CAN Error Frame

Generating the CAN error frame requires transforming the signal from digital to analogue. This is achieved using the same CAN transceiver used for sampling, to generate the differential signals required for CAN High and CAN Low bus lines, as shown below in figure 3. The CAN error frame consists of six dominant bits (One signal at 1.5V and the other at 3.5V), and eight recessive bit (two signals both at 2.5V). The generated error frame is then tested to prove that it works as intended through injecting (transmitting through transceiver '0') in the middle of a CAN frame being transmitted. This leads to fully hindering reception of malicious frame.

*Figure 4: CAN error frame signal generation.*

### 3.6. Prevention of Malicious CAN Frames

Having the frame sampled/analysed, and the error frame signal generated, CAN error frame is injected successfully to kill a CAN frame. This technique does not allow the CAN frame reaching the target CAN node. Then, CAN controller, of the attacker, repeatedly tries resending. This continues till switching to Bus-Off state at the end as shown below in figure 5. The attacker can no longer send any frames on the network. It requires hardware reset to reconnect.

*Figure 5: Killing CAN frame.*

All reliable message IDs are configured initially. The attacker is identified once the message ID is processed in the RbT node. All consequent malicious frames are detected and prevented until the attacker CAN node state is Bus-Off. Output is shown in figure 6.

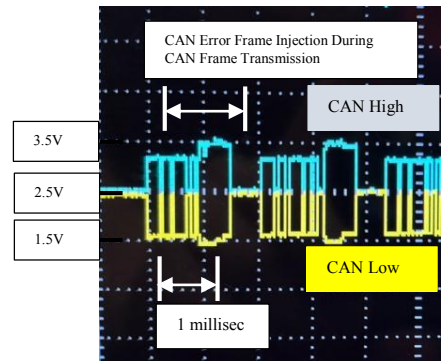

*Figure 6: Attacker CAN node in Bus-Off state.*

### 3.7. Time Measurements

Since the CAN bit rate is 10 Kbps, sampling each bit of the CAN frame is done once each 100 microsec. Sampling needs less than 1% of the software overhead.

The CAN error frame must be injected before acknowledgment bits. So, using TM4C 1294 EVB running at 80Mhz, max slack time is 1.6 microsec. for data-section-based security algorithm. On the other hand, 100 microsec. is available for CRC based algorithm.

### 4. CONCLUSION

This paper aims at solving cyber security vulnerability of CAN network. It proposes a new approach that detects a malicious CAN frame and resets the attacker ECU. This prevents the attacker frame from reaching the receiver CAN node. Consequently, the attacker node switches to Bus-Off state. CAN bus monitoring tool is implemented to acquire and present real time results. Experimental results show the efficiency of the proposed technique.

### 5. FUTURE WORK

Future work will focus on applying advanced cryptography algorithms during the slack time available.